\documentclass[a4paper,11pt]{article}
\pdfoutput=1
\usepackage{jcappub}
\usepackage{threeparttable}
\usepackage[T1]{fontenc}
\usepackage{color}

\DeclareFontFamily{OT1}{rsfs}{}
\DeclareFontShape{OT1}{rsfs}{m}{n}{<-7> rsfs5
    <7-10> rsfs7 <10-> rsfs10}{}
\DeclareMathAlphabet{\scr}{OT1}{rsfs}{m}{n}

\newcommand{\Mp}{M_\text{Pl}}
\newcommand{\M}{M_{\star}}
\newcommand{\di}{\partial}
\newcommand{\eff}{\text{eff}}

\renewcommand{\k}{\mathbf{k}}

\newcommand{\ENTRY}[1]{$\displaystyle {#1} $}

\title{Frustration of resonant preheating by exotic kinetic terms}

\author{Shohreh Rahmati}
\author{and Sanjeev S.\ Seahra}
\affiliation{Department of Mathematics and Statistics, University of New Brunswick \\ Fredericton, NB, Canada E3B 5A3}
\emailAdd{srahmati@unb.ca}
\emailAdd{sseahra@unb.ca}

\abstract{We study the effects of exotic kinetic terms on parametric resonance during the preheating epoch of the early universe.  Specifically, we consider modifications to the action of ordinary matter fields motivated by generalized uncertainty principles, polymer quantization, as well as Dirac-Born-Infeld and k-essence models.  To leading order in an ``exotic physics'' scale, the equations of motion derived from each of these models have the same algebraic form involving a nonlinear self-interaction in the matter sector.  Neglecting spatial dependence, we show that the nonlinearity effectively shuts down the parametric resonance after a finite time period.  We find numeric evidence that the frustration of parametric resonance persists to spatially inhomogenous matter in $(1+1)$--dimensions.}

\begin{document}
\maketitle
\flushbottom

\section{Introduction}

Motivated by dynamical dark energy models or theories of quantum gravity, many authors have considered the possibility that kinetic terms appearing in matter actions get modified at high or low energies.  In a cosmological context, exotic kinetic terms are often applied to scalar fields via direct modification of the covariant Lagrangian; examples of this include Dirac-Born-Infeld (DBI) inflation \cite{Silverstein:2003hf,Alishahiha:2004eh}, k-essence \cite{ArmendarizPicon:1999rj}, ghost condensates \cite{ArkaniHamed:2003uy}, and Galileon cosmology \cite{Nicolis:2008in}.  On the other hand, some approaches to quantum gravity suggest that conventional Schr\"odinger quantization is replaced with alternative quantizations at high energy, resulting in non-trivial semi-classical corrections to the kinetic part of scalar field Hamiltonians (among other effects).  Various classes of novel quantization schemes considered in the literature; including models involving generalized uncertainty principles (GUPs) or modified Heisenberg algebras \cite{Kempf:1994su,Hassan:2002qk,Das:2008kaa,Battisti:2009zzb,Chemissany:2011nq,Husain:2012im,Husain:2013zda,Day:2013gwa,Faizal:2014mfa,Faizal:2014pia,Faizal:2014rha}, and polymer quantization \cite{Ashtekar:2002sn,Ashtekar:2002vh,Husain:2003ry,Fredenhagen:2006wp,Corichi:2007tf,Husain:2007bj,Velhinho:2007gg,Hossain:2009vd,Hossain:2009ru,Hossain:2010eb,Hossain:2010wy,Seahra:2012un}.

The effects of exotic kinetic terms on the evolution of the cosmological background, the growth of classical perturbations, and the generation of primordial fluctuations during inflation have been extensively studied \cite{Clifton:2011jh}.   Such modifications have also been applied to the preheating phenomenon at the end of inflation, which is a mechanism to transfer the energy stored in the inflaton $\phi$ to ordinary matter \cite{Kofman:1997yn,Bassett:2005xm,Allahverdi:2010xz}.  This process relies on the assumptions that $\phi$ becomes oscillatory at the end of inflation (which is true in many models), and that there exists some simple coupling between $\phi$ and ordinary matter (usually modelled by an effective scalar degree of freedom $\chi$ called the ``reheaton'').  Oscillations in $\phi$ induce periodic variations in the effective mass of $\chi$, which in turn can cause the amplitude of $\chi$ to increase exponentially.  This is known as ``parametric resonance'' \cite{landau1976mechanics}.  Upon quantization, the exponential divergence of classical solutions for $\chi$ can be interpreted as ``explosive'' particle production of ordinary matter.

To incorporate exotic kinetic terms into this scenario, one can modify the action of the inflaton, the reheaton, or both.  Application of the DBI modification to the inflaton's kinetic term has been studied in references \cite{Davis:2009wg,Bouatta:2010bp,Karouby:2011xs,Child:2013ria,Zhang:2013asa}.  A particular type of scalar-tensor type modification to the reheaton's kinetic term was considered in \cite{Lachapelle:2008sy}.  The effects of GUPs on the reheaton dynamics were studied in reference \cite{Chemissany:2011nq}, where it was assumed that the exotic physics effects were extremely small.  The implications of a DBI kinetic term for the reheaton was recently analyzed in \cite{Underwood:2013pwa}.

The purpose of this paper is to systematically extend the analyses of \cite{Chemissany:2011nq} and \cite{Underwood:2013pwa} to several different types of models and choices of parameters.\footnote{We note that \cite{Underwood:2013pwa} appeared while this manuscript was in preparation, hence there is some overlap with our analysis of DBI modifications to reheaton dynamics in the homogeneous case.  In particular, our conclusion that DBI kinetic terms tend to frustrate parametric resonance agrees with \cite{Underwood:2013pwa}, but our analysis methods are somewhat different.  Furthermore, we consider a number of additional models in this paper.}  In section \ref{sec:models}, we review the basic equations governing preheating when kinetic terms are of the standard form, and then present modified reheaton equations of motion from GUP, polymer quantization, DBI, and k-essence models.  All of these models involve an ``exotic physics'' energy scale $\M$ which determines when kinetic terms differ significantly from the canonical choice.  Working to leading order in $\M^{-1}$ and in the long-wavelength limit, we demonstrate that the reheaton equation of motion for all these models takes the same form; that of a damped Mathieu-type equation modified by the addition of a nonlinear self-interaction.  For most models, the nonlinearity takes the form of a cubic polynomial in the reheaton amplitude and its time derivative.

In section \ref{sec:multiple scales}, we analyze this equation using a multiple scales approximation to describe near-resonance dynamics in terms of an autonomous dynamical system of ordinary differential equations.  Neglecting cosmological expansion, we find the fixed points of the system and classify their stability in section \ref{sec:non-expanding}.  We determine that the exponentially growing solutions familiar from canonical preheating theory are transformed into  bounded periodic solutions when the nonlinearity is present.  That is, the reheaton self-interactions implied by the models considered in this paper tend to frustrate parametric resonance, and hence curtail explosive particle production.  We confirm our analytic results by performing numeric simulations.  In section \ref{sec:expanding}, we re-analyze the equation of motion when the Hubble parameter is small but nonzero.  In this case, the dynamical system governing the reheaton evolution exhibits a rich bifurcation structure that we investigate in some detail.  However, the main conclusions from the non-expanding case are unaltered by the inclusion of a small Hubble parameter.

An open question is: do the results we obtained in section \ref{sec:multiple scales} for spatially homogeneous modes generalize to the inhomogeneous case?  In section \ref{sec:inhomogeneous}, we conduct numeric simulations of a spatially inhomogeneous reheaton in $(1+1)$--dimensions assuming the polymer quantization kinetic term.  We find that again, the nonlinearity appears to frustrate the parametric resonance effect. However, our results in this case are far less robust than in section \ref{sec:multiple scales}; the analysis is limited to numeric simulations and we do not attempt to make an exhaustive search of parameter space.  Nevertheless, we feel that the results of this section provide motivation for further work.

Section \ref{sec:conclusions} is reserved for our conclusions.

\section{Preheating models with exotic kinetic terms}\label{sec:models}

In this section, we review the basics of the preheating mechanism assuming canonical kinetic terms and then derive modified reheaton equations of motion for GUP, polymer quantization, DBI, and k-essence models.  To leading order in an ``exotic physics'' scale $\M^{-1}$, we find that the reheaton's equation of motion is the same for each model in the spatially homogeneous limit, up to the choice of three dimensionless parameters.

\subsection{Background evolution and the canonical reheaton equation of motion}

We consider a theory of two interacting scalar fields $\phi$ and $\chi$ in the presence of gravity:
\begin{equation}
	S[\phi,\chi,g] = \int d^{4}x \sqrt{-g} \left[ \frac{1}{2} \Mp^{2} R - \frac{1}{2} \nabla^{\mu} \phi \, \nabla_{\mu} \phi - \frac{1}{2} m_{\phi}^{2} \phi^{2} + \mathcal{L}_m (\chi,\phi)  \right].
\end{equation}
In this expression $\phi$ is taken to be the inflaton field, while the reheaton $\chi$ represents some ``matter'' degree of freedom.  We assume that the matter Lagrangian $\mathcal{L}_{m}$ contains an interaction term:
\begin{equation}
	-\frac{1}{2} g \phi^{2} \chi^{2} \subset \mathcal{L}_{m}.
\end{equation}
Furthermore, we take the line element to be of the Friedmann-Robertson-Walker form:
\begin{equation}
	ds^{2} = -dt^{2} + a^{2}(t) d\mathbf{x}^{2}.
\end{equation}
During inflation, the homogenous mode of the inflaton is taken to be much larger than the matter field, so we can initially neglect $\mathcal{L}_{m}$.  Under this assumption, the equations of motion are:
\begin{subequations}
\begin{gather}
\ddot{\phi} + 3H\dot\phi+m_{\phi}^{2}\phi = 0, \\ H^{2} = \tfrac{1}{6} \Mp^{-2} ( \dot\phi^{2} + m_{\phi}^{2} \phi^{2} ),
\end{gather}
\end{subequations}
where $H = \dot{a}/a$ is the usual Hubble parameter.

The basic features of solutions of these equations are well-known. If the initial field value is super-Planckian, $\phi \gtrsim \Mp$ then the cosmological dynamics will consist of a transient period followed by an era of slow-roll inflation where the field amplitude decreases slowly.  When $\phi$ decreases below $\Mp$, inflation ends and the inflaton oscillates about the minimum of the potential $\frac{1}{2} m^{2}_{\phi} \phi^{2}$.  During this epoch, we have
\begin{equation}\label{eq:oscillating phi}
	\phi(t) \approx \Phi(t) \cos (m_{\phi}t),
\end{equation}
where $\Phi(t) \propto 1/t$ is a function that varies slowly over the Hubble time $H^{-1}$.

The phenomenon of preheating takes place in the era where the approximation (\ref{eq:oscillating phi}) holds.  We consider timescales $\lesssim H^{-1}$, which means we can treat $\Phi(t)$ as constant.  The dynamics of the matter field $\chi$ is then governed by the reduced action
\begin{equation}
	S_m[\chi] = \int d^{4}x \,  a^{3} \mathcal{L}_{m}(\chi,\phi ).
\end{equation}
The standard choice of matter Lagrangian in the preheating literature is
\begin{equation}
	\mathcal{L}_{m} = - \frac{1}{2} \nabla^{\mu} \chi \, \nabla_{\mu} \chi - \frac{1}{2} m_{\chi}^{2} \chi^{2} - \frac{1}{2} g \phi^{2} \chi^{2},
\end{equation}
which gives rise to the $\chi$ equation of motion
\begin{equation}
	\ddot \chi  + 3H \dot\chi +\left(- \frac{\nabla^{2}}{a^{2}} + m_{\chi}^{2} + g\phi^{2}\right)\chi  = 0, \quad \nabla^{2} = \di_{x}^{2}+\di_{y}^{2}+\di_{z}^{2}.
\end{equation}
During the preheating epoch this becomes:
\begin{equation}
	\ddot \chi  + 3H \dot\chi - \frac{\nabla^{2}}{a^{2}} \chi + [m_{\chi}^{2} + g\Phi^{2} \cos^{2}(m_{\phi}t) ]  \chi = 0.
\end{equation}
We see that the effective mass of the $\chi$ field is harmonically modulated, which can exponentially amplify the $\chi$ field's amplitude via parametric resonance under certain circumstances.  The main goal of this paper is to study what happens to parametric resonance when the matter involves the non-standard features described in the following subsections.

\subsection{Generalized uncertainty principles (GUPs)}

One possible consequence of quantum gravity is that the Heisenberg algebra of quantum mechanics gets modified at high energies.  Many authors have considered the phenomenological implications of this idea by assuming the quantum commutator between configuration variables $\{q_{i}\}_{i=1}^{N}$ and their conjugate momenta $\{p_{j}\}_{j=1}^{N}$ is deformed to
\begin{equation}
	[\hat{q}_{i},\hat{p}_{j}] = if_{ij}(\hat{p}_{1},\hat{p}_{2} \cdots \hat{p}_{N}).
\end{equation}
Here, we consider the effects of such modified commutators on the equations of motion of the matter field $\chi$ during preheating.  As described in detail in reference \cite{Chemissany:2011nq}, specific choices of $f_{ij}$ give rise to various types of quantum corrections to the dynamics, which can be encoded in an effective action
\begin{equation}
	S_{m}^{\eff}[\chi] = \int d^{4}x \left[ p_{\chi}\dot{\chi} - \mathcal{H}_{m}^{\eff}(\chi,p_{\chi}) \right].
\end{equation}
For a particular class of $f_{ij}$, the effective Hamiltonian density takes the form:\footnote{This is the effective Hamiltonian studied by \cite{Chemissany:2011nq} generalized to an FRW background.}
\begin{equation}
	\mathcal{H}_{m}^{\eff}(\chi,p_{\chi}) =  \frac{p_{\chi}^{2}}{2a^{3}} \left[ 1 + \kappa \left( \frac{p_{\chi}}{\M^{2}a^{3}} \right)^{n} \right] + \frac{1}{2} a(\mathbf{\nabla} \chi)^{2} +  \frac{1}{2} a^{3} \left[ m_{\chi}^{2}  + g \phi^{2} \right] \chi^{2},
\end{equation}
where $\kappa = \pm 1$, $n$ is a integer, and $\M$ is mass space that determines when deformed algebra effects are important.  The particular values of $n$ and $\kappa$ are determined by the exact form of the generalized uncertainty relation; two common choices are $n = 1$ and $n=2$.  Hamilton's equations are
\begin{equation}\label{eq:Hamilton's equations}
	\dot{\chi} = \frac{\di \mathcal{H}_{m}^{\text{eff}}}{\di p_{\chi}}, \quad \dot{p}_{\chi} = -\frac{\di \mathcal{H}_{m}^{\text{eff}}}{\di \chi} + \di_{i} \frac{\di \mathcal{H}_{m}^{\text{eff}}}{\di (\di_{i}\chi)},
\end{equation}
which yield
\begin{equation}
	\dot\chi= \frac{p_{\chi}}{a^{3} }\left[ 1 + \kappa\left(\frac{n+2}{2} \right) \left(\frac{p_{\chi}}{\M^{2}a^{3}}\right)^{n} \right], \quad \dot{p}_{\chi} = -a^{3}(m_{\chi}^{2} + g \phi^{2}) \chi + a\nabla^{2}\chi.
\end{equation}
These can be re-arranged into a single wave equation for $\chi$:
\begin{multline}
	\ddot \chi  + 3H \dot\chi +  \left(- \frac{\nabla^{2}}{a^{2}} + m_{\chi}^{2} + g\phi^{2}\right)\chi  = \\ - \frac{\kappa}{2}(n+2)  \left( \frac{\dot\chi}{\M^{2}} \right)^{n} \left[ (n+1)  \left(- \frac{\nabla^{2}}{a^{2}} + m_{\chi}^{2} + g\phi^{2}\right)\chi +3nH\dot\chi \right] + \mathcal{O} \left( \frac{\dot\chi^{2n}}{\M^{4n}}\right).
\end{multline}
Here, we have implicitly assumed that $\M$ is a large mass scale.

We now examine this equation in the long wavelength limit where spatial derivatives can be ignored.  We also drop $\mathcal{O}(\dot{\chi}^{2n}/\M^{4n})$ terms, and assume that we are in the preheating phase $\phi = \Phi\cos(m_{\phi}t)$.  It is useful to define a characteristic frequency:
\begin{equation}\label{eq:characteristic frequency}
	\omega_{0}^{2} = m_{\chi}^{2} + \frac{1}{2} g \Phi^{2},
\end{equation}
with which we can define a dimensionless time and scalar field variable:
\begin{equation}\label{eq:dimensionless}
	T = \omega_{0} t, \quad X = \frac{\chi}{\omega_{0}}.
\end{equation}
In terms of these quantities, the equation of motion reads
\begin{equation}
	\frac{d^{2}X}{dT^{2}} + 3\Theta \frac{dX}{dT} + (1 + h \cos \Omega T)X - \mu \left( \frac{dX}{dT} \right)^{n} \left[ (1 + h \cos \Omega T)X + \frac{3n\Theta}{n+1}  \frac{dX}{dT} \right] = 0,
\end{equation}
where we have defined the dimensionless parameters:
\begin{equation}\label{eq:h and Theta def}
	h = \frac{g\Phi^{2}}{2\omega_{0}^{2}}, \quad \Omega = \frac{2m_{\phi}}{\omega_{0}} = \frac{2 m_{\phi}\sqrt{1-h}}{m_{\chi}}, \quad \Theta = \frac{H}{\omega_{0}},
\end{equation}
in addition to the dimensionless nonlinear coupling
\begin{equation}
         \mu = - \frac{\kappa(n+1)(n+2)}{2} \left( \frac{\omega_{0}}{\M} \right)^{2n}.
\end{equation}

\subsection{Polymer quantization}

A different phenomenological implication of quantum gravity may imply that conventional Schr\"odinger quantization needs to be replaced by an alternate scheme at high energy: namely polymer quantization \cite{Ashtekar:2002sn}.  In the semi-classical approximation, it is possible to encode polymer quantization effects in a effective Hamiltonian of the form \cite{Hossain:2009vd}:
\begin{equation}
	\mathcal{H}_{m}^{\eff}(\chi,p_{\chi}) = \frac{\M^{4}a^{3}}{2} \sin^{2}\left( \frac{p_{\chi}}{\M^{2} a^{3}} \right) + \frac{1}{2} a(\mathbf{\nabla} \chi)^{2} +  \frac{1}{2} a^{3}\left[ m_{\chi}^{2}  + g \phi^{2} \right] \chi^{2},
\end{equation}
where $\M$ is a mass scale that defines when the polymer quantization effects are important.  In this case, Hamilton's equations (\ref{eq:Hamilton's equations}) yield
\begin{equation}
	\dot\chi = \frac{\M^{2}}{2} \sin\left( \frac{2p_{\chi}}{\M^{2}a^{3}} \right), \quad  \dot{p}_{\chi} = -a^{3}(m_{\chi}^{2} + g \phi^{2}) \chi +  a\nabla^{2}\chi.
\end{equation}
These can be re-arranged into a single wave equation for $\chi$:
\begin{equation}\label{eq:polymer EOM}
	\ddot\chi + \left( 1 - \frac{4\dot\chi^{2}}{\M^{4}}  \right)^{1/2} \left[ \left( - \frac{\nabla^{2}}{a^{2}} +m_{\chi}^{2} + g \phi^{2}\right)\chi + \frac{3}{2} \M^{2} H \arcsin \left( \frac{2\dot\chi}{\M^{2}} \right) \right] = 0.
\end{equation}

Let us examine this equation in the long-wavelength limit where spatial derivatives can be ignored and in the preheating epoch when $\phi = \Phi\cos(m_{\phi}t)$.  Defining a characteristic frequency $\omega_{0}$, dimensionless time $T$, and dimensionless field amplitude $X$ as in equations (\ref{eq:characteristic frequency}) and (\ref{eq:dimensionless}), respectively, we obtain that:
\begin{equation}
	\frac{d^{2}X}{dT^{2}} + \left[ 1- 2\mu \left( \frac{dX}{dT} \right)^{2} \right]^{1/2} \left[ (1 + h \cos \Omega T) X  + \frac{3}{\sqrt{2\mu}} \Theta \arcsin \left( \sqrt{2\mu} \frac{dX}{dT} \right) \right] = 0.
\end{equation}
Here, we have defined the dimensionless parameters $h$, $\Omega$ and $\Theta$ as in (\ref{eq:h and Theta def}) and the nonlinear coupling as
\begin{equation}
	\mu = 2 \left( \frac{\omega_{0}}{\M} \right)^{4}.
\end{equation}

\subsection{Dirac-Born-Infeld (DBI)}

If we assume that the ``matter'' degree of freedom is actually a moduli field that measures the position of a D-brane in a higher dimensional manifold, its low energy effective Lagrangian will have a kinetic term of the Dirac-Born-Infeld (DBI) form \cite{Silverstein:2003hf,Alishahiha:2004eh}:
\begin{equation}\label{eq:DBI Lagrangian}
	\mathcal{L}_{m}^{\eff} = \Lambda^{4}(\chi) \left[ 1-\left( 1+\frac{\nabla^{\mu}\chi \, \nabla_{\mu}\chi }{\Lambda^{4}(\chi)} \right)^{1/2} \right]- \frac{1}{2} m_{\chi}^{2} \chi^{2} - \frac{1}{2} g \phi^{2} \chi^{2}.
\end{equation}
For simplicity, we take the ``warp factor'' to be equal to a constant mass scale:
\begin{equation}
	\Lambda(\chi) = \M.
\end{equation}
The Euler-Lagrange equations are
\begin{equation}\label{eq:Euler-Lagrange}
	\nabla_{\mu} \left[ \frac{\di\mathcal{L}_{m}^{\text{eff}}}{\di(\nabla_{\mu} \chi)} \right] = \frac{\di\mathcal{L}_{m}^{\text{eff}}}{\di \chi}.
\end{equation}
This gives the equation of motion
\begin{equation}
	-\Box\chi + \frac{(g^{\rho\mu}g^{\lambda\nu} - g^{\rho\lambda}g^{\mu\nu}) \nabla_{\rho}\chi \, \nabla_{\lambda}\chi \, \nabla_{\mu}\nabla_{\nu}\chi }{\M^{4}} + \left(1 + \frac{\nabla^{\mu}\chi \, \nabla_{\mu}\chi }{\M^{4}} \right)^{3/2} (m_{\chi}^{2} + g \phi^{2}) \chi^{2} = 0.
\end{equation}

Let us examine this equation in the long-wavelength limit where spatial derivatives can be ignored and in the preheating epoch when $\phi = \Phi\cos(m_{\phi}t)$.  Defining a characteristic frequency $\omega_{0}$, dimensionless time $T$, and dimensionless reheaton field amplitude $X$ as in equations (\ref{eq:characteristic frequency}) and (\ref{eq:dimensionless}), respectively, we obtain that:
\begin{equation}
	\frac{d^{2}X}{dT^{2}} + 3\Theta\frac{dX}{dT}+ \left[ 1- \frac{2\mu}{3} \left( \frac{dX}{dT} \right)^{2} \right]^{3/2} (1 + h \cos \Omega T) X - 2\mu\Theta \left( \frac{dX}{dT} \right)^{3} = 0.
\end{equation}
Here, we have defined the dimensionless parameters $h$ and $\Omega$ as in (\ref{eq:h and Theta def}) and the nonlinear coupling as
\begin{equation}
	\mu = \frac{3}{2} \left( \frac{\omega_{0}}{\M} \right)^{4}.
\end{equation}

\subsection{k-essence}

The DBI Lagrangian (\ref{eq:DBI Lagrangian}) is actually a particular type of k-essence model \cite{ArmendarizPicon:1999rj} where the kinetic term for a scalar field has a nonstandard form.  A sub-category of k-essence models that encompasses the DBI action is given by
\begin{equation}
	\mathcal{L}_{m}^{\eff} = \frac{\M^{4}}{2} P(\zeta) - \frac{1}{2} m_{\chi}^{2} \chi^{2} - \frac{1}{2} g \phi^{2} \chi^{2}, \quad \zeta = -\frac{\nabla^{\mu}\chi \nabla_{\mu} \chi}{\M^{4}},
\end{equation}
where $P$ is a dimensionless function of one variable.  In this paper, we wish to restrict ourselves to k-essence models that reproduce ordinary results when $\M \rightarrow \infty$.  Hence, we assume that the series expansion of $P$ about 0 is of the form:
\begin{equation}
	P(\zeta) = \zeta + \frac{\beta}{k+1} \zeta^{k+1} + \mathcal{O}(\zeta^{k+2}),
\end{equation}
where $k \ge 1$ is an integer and $\beta$ is a number.  Applying the Euler-Lagrange equations to this Lagrangian yields:
\begin{equation}
	-P'(\zeta) \Box \chi + \frac{2}{\M^{4}} P''(\zeta) \nabla^{\mu} \chi \, \nabla^{\nu} \chi \nabla_{\mu} \nabla_{\nu} \chi + (m_{\chi}^{2} + g \phi^{2}) \chi^{2} = 0.
\end{equation}
Specializing to the long-wavelength limit where spatial derivatives are dropped, we obtain:
\begin{equation}
	 \ddot{\chi} +3H\dot\chi +  (m_{\chi}^{2} + g \phi^{2}) \chi - \beta \left(\frac{\dot\chi}{\M^{2}} \right)^{2k}   \left[   6Hk \dot\chi  + (1+2k)  (m_{\chi}^{2} + g \phi^{2}) \chi  \right]=\mathcal{O} \left( \frac{\dot\chi^{2(k+1)}}{\M^{4(k+1)}} \right).
\end{equation}
In the preheating epoch, we have $\phi = \Phi\cos(m_{\phi}t)$.  Dropping higher order terms, and defining a characteristic frequency $\omega_{0}$, dimensionless time $T$, and dimensionless field amplitude $X$ as in equations (\ref{eq:characteristic frequency}) and (\ref{eq:dimensionless}), respectively, we obtain that:
\begin{equation}
	\frac{d^{2}X}{dT^{2}} +3\Theta\frac{dX}{dT} +  (1 + h \cos \Omega T)X  - \mu \left( \frac{dX}{dT} \right)^{2k} \left[ (1 + h \cos \Omega T) X + \frac{6k\Theta}{1+2k} \frac{dX}{dT} \right]= 0.
\end{equation}
Also, we have defined the dimensionless parameters $h$, $\Omega$ and $\Theta$ as in (\ref{eq:h and Theta def}) and the nonlinear coupling as
\begin{equation}
	\mu = \beta(1+2k) \left( \frac{\omega_{0}}{\M} \right)^{4k}.
\end{equation}

\subsection{Summary}

To leading order in the small parameter $\mu$, all the matter equations of motion derived in this section can be written as
\begin{equation}\label{eq:general ode}
	\frac{d^{2}X}{dT^{2}} +3\Theta\frac{dX}{dT} +  (1 + h \cos \Omega T)X = \mu \left( \frac{dX}{dT} \right)^{n} \left[ (1 + h \cos \Omega T) X + q \Theta \frac{dX}{dT} \right],	
\end{equation}
where $h$, $\Omega$ and $\Theta$ are given by (\ref{eq:h and Theta def}) while $\mu$, $q$ and $n$ vary from model to model, and are summarized in table \ref{tab:summary}.
\begin{table}
\begin{center}
\begin{tabular}{ccccc}
\hline\hline model & $\mu$ & $q$ & $n$ & parameters \\
\hline GUPs & $- \frac{1}{2}\kappa(n+1)(n+2) \left( \omega_{0}/\M \right)^{2n}$ & $3n/(n+1)$ & $1,2$ & $\kappa = \pm 1$ \\
polymer quantization & $2 \left( \omega_{0}/\M \right)^{4}$ & 2 & 2 & n/a \\
Dirac-Born-Infled & $\frac{3}{2} \left(  \omega_{0}/\M \right)^{4}$ & 2 & 2 & n/a \\
k-essence & $\beta(1+2k) \left(  \omega_{0}/\M \right)^{4k}$ & $6k/(1+2k)$ & $2k$ & $\beta \in \mathbb{R}$, $k \in \mathbb{Z}^{+}$ \\
\hline
\end{tabular}
\end{center}
\caption{Summary of the values of the parameters appearing in the generic equation of motion (\ref{eq:general ode}) for the various models considered in this paper.  In all cases, $\M$ is an energy scale that indicates the threshold for exotic physics and $\omega_{0}$ is the natural frequency of the $\chi$ field in the limit $\mu \rightarrow 0$, $h\rightarrow 0$ and $\Theta \rightarrow 0$.}\label{tab:summary}
\end{table}

\section{Multiple scales analysis}\label{sec:multiple scales}

We now consider the solutions of (\ref{eq:general ode}) when the nonlinear coupling parameter and the inflaton-matter coupling are small; that is, $|\mu| \ll 1$ and $|h| \ll 1$.  To simplify the analysis, we will concentrate on the case $n=q=2$, which is common to all of the models summarized in table \ref{tab:summary}.   We also tune the dimensionless frequency of the inflaton $\Omega$ to be close to the optimal value for parametric resonance.

\subsection{Non-expanding background}\label{sec:non-expanding}

First, we examine (\ref{eq:general ode}) in the case where the expansion of the background is negligible; i.e., in the limit $\Theta \rightarrow 0$.  The equation of motion reduces to:
\begin{equation}\label{eq:general ode 2}
	\frac{d^{2}X}{dT^{2}} + (1 + h \cos \Omega T) X = \mu (1 + h \cos \Omega T) \left( \frac{dX}{dT} \right)^{2}X  .	
\end{equation}
If we set $\mu = 0$, we obtain the well known Mathieu equation \cite{abramowitz+stegun}:
\begin{equation}
	\frac{d^{2}X}{dT^{2}} + (1 + h \cos \Omega T) X = 0.	
\end{equation}
This differential equation exhibits the parametric resonance phenomena when $\Omega \approx 2/N$ with $N = 1,2,3 \ldots$ \cite{landau1976mechanics}.  The fundamental resonance $N = 1$ is the ``strongest'' in the sense that it is the fastest growing instability of the equation.  For the fundamental mode ($\Omega \approx 2$), an approximate solution is
\begin{equation}\label{eq:approx solution}
	X(T) \approx a(T) \cos \left( \tfrac{1}{2} \Omega T \right) + b(T) \sin \left( \tfrac{1}{2} \Omega T \right), \quad a(T) = a_{0} e^{sT}, \quad b(T) = b_{0} e^{sT},
\end{equation}
where $s > 0$, $a_{0}$ and $b_{0}$ are constants.

We seek an approximate solution of the same form as (\ref{eq:approx solution}) when $\mu \ne 0$ in (\ref{eq:general ode 2}).  We restrict our analysis to $n=2$.  We adopt the ansatz:
\begin{equation}\label{eq:ansatz}
	X(T) = a(T) \cos \left( \tfrac{1}{2} \Omega T \right) + b(T) \sin \left( \tfrac{1}{2} \Omega T \right).
\end{equation}
We work under the assumption that $a(T)$ and $b(T)$ are slowly-varying compared to the oscillations of the trigonometric functions, which means that $\sqrt{a^{2} + b^{2}}$ can be interpreted as the overall amplitude $X(T)$.  Also, we assume that $\Omega$ is tuned close to the fundamental resonance; deviations of $\Omega$ from 2 are parametrized by $\epsilon$, defined as
\begin{equation}
	\epsilon = \frac{\Omega^{2}-4}{2}, \quad |\epsilon| \ll 1.
\end{equation}
We substitute (\ref{eq:ansatz}) into (\ref{eq:general ode 2}) and convert all the trigonometric functions into complex exponentials.  We then neglect:
\begin{itemize}
\item rapidly varying terms proportional to $e^{\pm \frac{3}{2}i\Omega T}$ and $e^{\pm \frac{5}{2}i\Omega T}$; \
\item second time derivatives of the amplitudes: $\ddot{a}$ and $\ddot{b}$; and
\item products of $\dot{a}$ and $\dot{b}$ with $\mu$.
\end{itemize}
Setting the coefficients of $e^{\pm \frac{1}{2} i \Omega T}$ equal to zero in the resulting expression yields a set of two coupled first order equations for $a$ and $b$:
\begin{subequations}\label{eq:dynamical system}
\begin{align}
	\frac{da}{dz} & = -\frac{1}{2} (\epsilon+h) b +\frac{1}{8} \mu b[2a^{2} h^{2} -(2-h)(a^{2}+b^{2})  ], \\
	\frac{db}{dz} & = +\frac{1}{2} (\epsilon-h) a - \frac{1}{8} \mu a[2b^{2} h^{2} -(2+h)(a^{2}+b^{2})],
\end{align}
\end{subequations}
where $z = T/\Omega$.

Equations (\ref{eq:dynamical system}) are an autonomous 2-dimensional dynamical system of the form:
\begin{equation}
	\frac{d \textbf{a}}{dz} = \textbf{f}(\textbf{a}), \quad \textbf{a} = \left( \begin{array}{c} a \\ b \end{array} \right), \quad \textbf{f}(\textbf{a}) = \left( \begin{array}{c} f_{1}(a,b) \\ f_{2}(a,b) \end{array} \right).
\end{equation}
As usual, the qualitative behaviour of solutions of (\ref{eq:dynamical system}) is dictated by the nature of the system's fixed points \cite{wiggins2003introduction}; i.e.\ points $\mathbf{a}_{0} \in \mathbb{R}^{2}$ for which $\mathbf{F}(\mathbf{a}_{0}) = 0$.   For (\ref{eq:dynamical system}) there are 9 candidates for fixed points, as summarized in table \ref{tab:fixed points}.  Not all of these are in $\mathbb{R}^{2}$ for certain choices of $(\epsilon,h,\mu)$.  For example, for small $h$ and $\epsilon$ we see that $\delta_{\pm,\pm}$ are located at
\begin{equation}
	a_{0}^{2} \approx \frac{2+\epsilon}{h\mu} , \quad b_{0}^{2} \approx - \frac{2+\epsilon}{h\mu} .
\end{equation}
Since $a_{0}^{2} \approx - b_{0}^{2}$, it is impossible for both $a_{0}$ and $b_{0}$ to be real.  Hence, the $\delta_{\pm,\pm}$ fixed points are not relevant for the dynamics of real solutions of (\ref{eq:dynamical system}) when $h$ and $\epsilon$ are small.  We will hence ignore them for most of the following analysis.
\begin{table}
\begin{center}
\begin{tabular}{cccc}
\hline\hline fixed point & \ENTRY{a_{0}^{2}} & \ENTRY{b_{0}^{2}} & squared eigenvalues $\lambda^{2}$ of Jacobian \\ \hline
$\Xi$ & \ENTRY{0 \vphantom{\frac{a}{b}}} & 0 & \ENTRY{\frac{1}{4}(h+\epsilon)(h-\epsilon)}  \\
$\alpha_{\pm}$ & \ENTRY{\frac{4(h-\epsilon)}{\mu(h+2)}} & 0 & \ENTRY{-h(h-\epsilon)+\mathcal{O}(h^{2},\epsilon^{2},h\epsilon)} \\
$\beta_{\pm}$ & 0 &  \ENTRY{\frac{4(h+\epsilon)}{\mu(h-2)}} & \ENTRY{-h(h+\epsilon)+\mathcal{O}(h^{2},\epsilon^{2},h\epsilon)} \\
$\delta_{\pm,\pm}$ & \ENTRY{ \frac{2(\epsilon h + h^{2} - \epsilon - 2)}{\mu h(h^{2}-2)} }  & \ENTRY{ \frac{2(\epsilon h - h^{2} + \epsilon + 2)}{\mu h(h^{2}-2)} } & not relevant for $|h|,|\epsilon| \ll 1$ \\ \hline\hline
\end{tabular}
\end{center}
\caption{The fixed points of the dynamical system (\ref{eq:dynamical system})}\label{tab:fixed points}
\end{table}

We can linearize (\ref{eq:dynamical system}) about each of the fixed points:
\begin{equation}
	\mathbf{a} = \mathbf{a}_{0} + \varepsilon \mathbf{a}_{1}, \quad \frac{d}{dz} \mathbf{a}_{1}  = J  \mathbf{a}_{1} + \mathcal{O}(\varepsilon), \quad J = \left( \begin{array}{cc} \di_{a} f_{1} & \di_{b}f_{1} \\ \di_{a}f_{2} & \di_{b} f_{2} \end{array} \right)_{\mathbf{a} = \mathbf{a}_{0}},
\end{equation}
where $J$ is the Jacobian matrix of $\mathbf{F}$ evaluated at the fixed point, and $\varepsilon$ is a small perturbation parameter.  The eigenvalues $\lambda$ of $J$ determine the qualitative behaviour of solutions near each fixed point, and are also listed in table \ref{tab:fixed points}.  Notice that $\lambda^{2}$ is real for each of $\Xi$, $\alpha_{\pm}$, and $\beta_{\pm}$, which implies that the eigenvalues are either purely real or purely imaginary.  Based on the signs of $a_{0}^{2}$, $b_{0}^{2}$, and $\lambda^{2}$ for given $(\epsilon,h,\mu)$ we can determine if a fixed point is in $\mathbb{R}^{2}$, and, if so, if it is a saddle point ($\lambda^{2}>0$) or it is a circle ($\lambda^{2}<0$).  The results of this classification procedure are given in table \ref{tab:classification}.
\begin{table}
\begin{minipage}{0.7\textwidth}
\begin{tabular}{clcccc}
\hline\hline
region & definition  & $\mu$ & $\Xi$ & $\alpha_{\pm}$ & $\beta_{\pm}$ \\ \hline
IA & $\{ (\epsilon,h) \, | \, 0 < h < \epsilon \}$ & + & circle & n/a & n/a \\
    & & $-$ & circle & saddle & circle \\
IB &  $\{ (\epsilon,h) \, | \, -\epsilon < h < 0 \}$ & + & circle & n/a & n/a \\
    & & $-$ & circle & circle  & saddle \\
IIA & $\{ (\epsilon,h) \, | \, |\epsilon| < h \}$ & + & saddle & circle & n/a \\
    & & $-$ & saddle & n/a & circle \\
IIB & $\{ (\epsilon,h) \, | \, h< - |\epsilon| \}$ & + & saddle & n/a & circle  \\
    & & $-$ & saddle & circle & n/a \\
IIIA & $\{ (\epsilon,h) \, | \, 0 < h < -\epsilon \}$ & + & circle & circle & saddle  \\
    & & $-$ & circle & n/a & n/a \\
IIIB & $\{ (\epsilon,h) \, | \, \epsilon < h < 0 \}$ & + & circle & saddle & circle  \\
    & & $-$ & circle & n/a & n/a \\
\hline\hline
\end{tabular}
\end{minipage}%
\begin{minipage}{0.3\textwidth}
\begin{center}
\includegraphics[width=\textwidth]{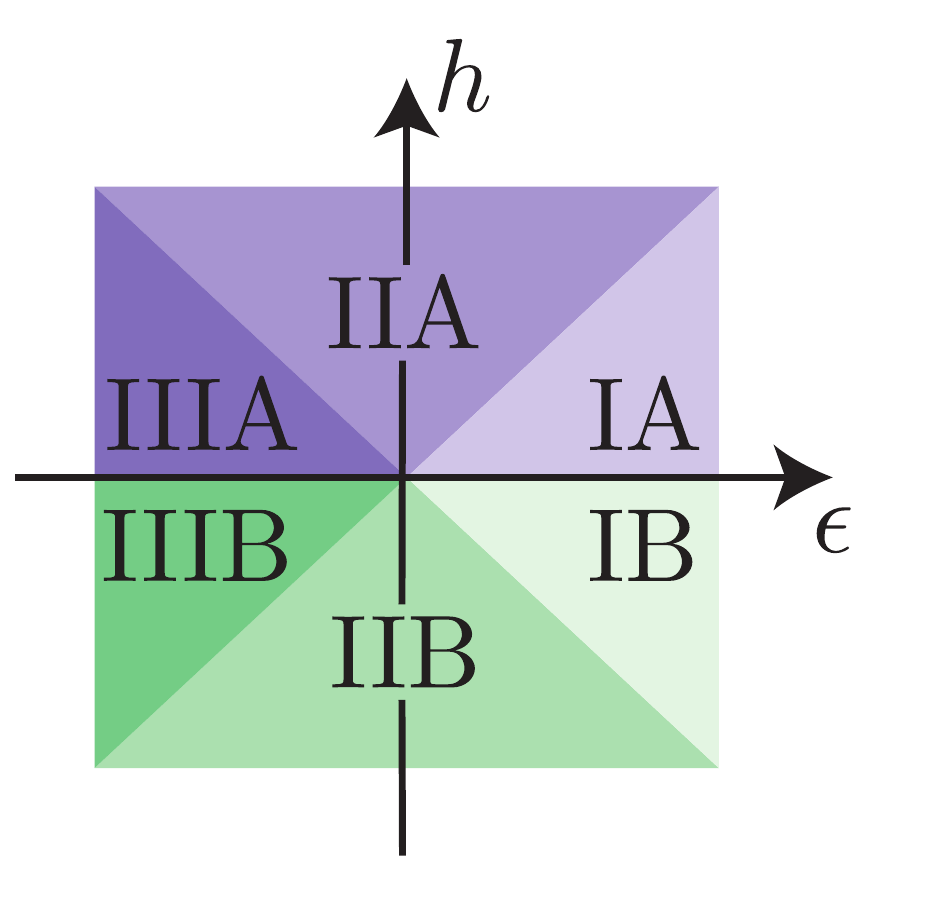}
\end{center}
\end{minipage}%
\caption{Classification of the fixed points of the dynamical system (\ref{eq:dynamical system}) based on values of the $(\epsilon,h,\mu)$ parameters.  The entry ``n/a'' indicates that the associated fixed point is not in $\mathbb{R}^{2}$ for that choice of parameters.  Note that the special cases $h = \pm \epsilon$ and $h=0$ are not included  in the table.}\label{tab:classification}
\end{table}

\begin{figure}
\begin{center}
\includegraphics[width=\textwidth]{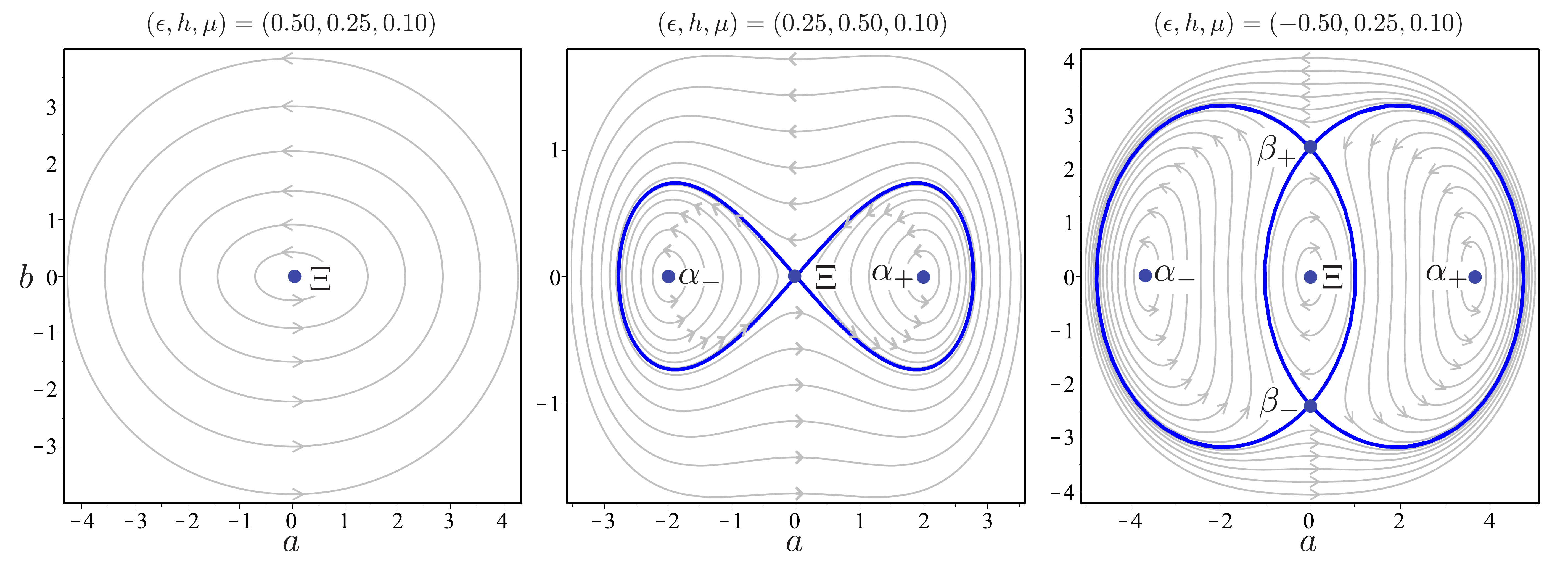}
\end{center}
\caption{Phase portraits of dynamics in regions IA (\emph{left}), IIA (\emph{centre}) and IIIA (\emph{right}) as obtained from the numerical solutions of the dynamical system (\ref{eq:dynamical system}).  The heavy blue lines are separatrices.  Portraits in regions IB, IIB or IIIB are qualitatively obtained by reflecting the above in the line $h=\epsilon$.}\label{fig:phase}
\end{figure}

Parametric resonance occurs when the vacuum fixed point $\Xi$ is unstable; i.e., it is a saddle as opposed to a circle.  The necessary conditions for this are independent of $\mu$:
\begin{equation}
	|\epsilon| < |h|.
\end{equation}
In this case, generic initial data with $|a| \ll 1$ and $|b| \ll 1$ will induce trajectories repulsed away from $a = b = 0$ along preferred directions in the $(a,b)$ plane.  In the absence of the nonlinear term (i.e., $\mu = 0$) these trajectories would approach infinity, indicating that the parametric resonance induces the $X$ field to grow without bound.  However, when $\mu \ne 0$ we find that there exist two additional circular fixed points on the $a$ or $b$ axes.  In the middle panel of figure (\ref{fig:phase}) one can see the effect of these extra fixed points is to ensure that trajectories repulsed from $\Xi$ do not wander off to infinity; rather, they settle into periodic orbits focussed around either $\Xi$ or one of the other fixed points.  That is, when $\mu$ is nonzero, the parametric resonance effect is frustrated.

\begin{figure}
\begin{center}
\includegraphics[width=\textwidth]{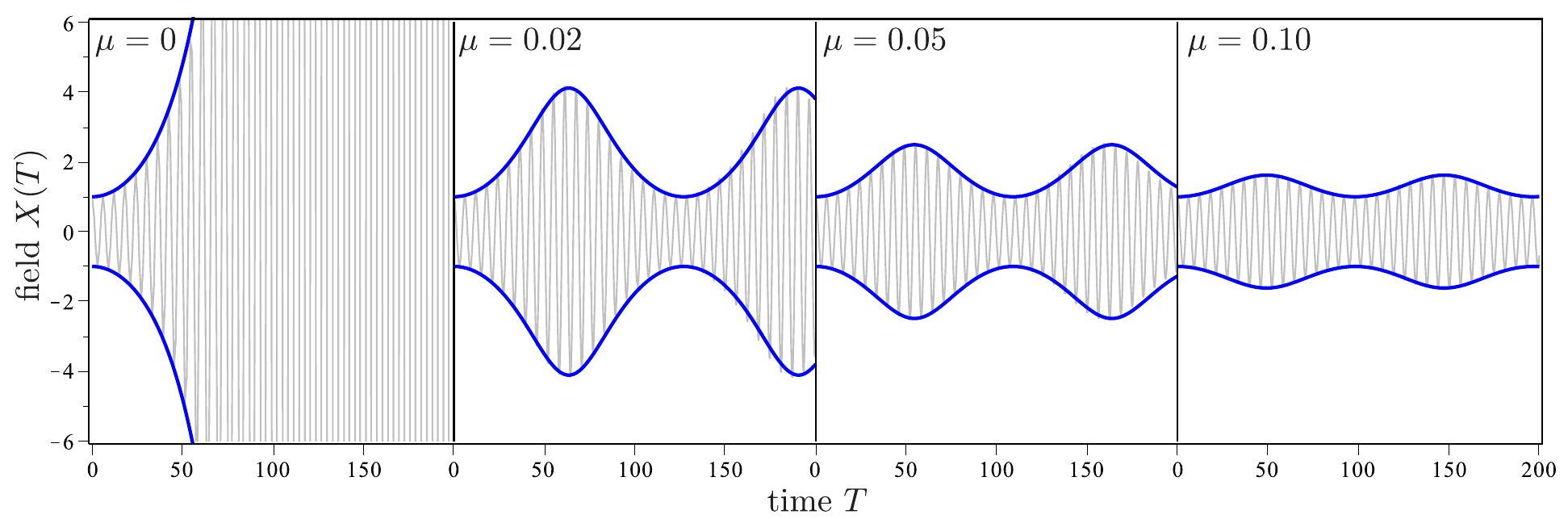}
\end{center}
\caption{Numeric simulations of the solutions of the equation motion (\ref{eq:general ode 2}) for $\epsilon = 0.1$ and $h=0.2$ (\emph{grey}).  Also shown (\emph{blue}) is the predicted amplitude $\sqrt{a^{2}+b^{2}}$ of oscillations of $X(T)$, as obtained from the numerical solution of the multiple scales equations (\ref{eq:dynamical system}).  We see fairly good agreement between the direct numerical solution and the multiple scales result.  We also see that increasing the magnitude of the nonlinearity $\mu$ tends to increasingly frustrate the parametric resonance effect.}\label{fig:simulations}
\end{figure}

This conclusion is clearly illustrated in figure \ref{fig:simulations}, where we plot the direct numerical solution of (\ref{eq:general ode 2}) as well as the associated prediction from the multiple scales analysis.  In this plot, we have selected values of $h$ and $\epsilon$ that would give rise to parametric resonance if $\mu = 0$.  Clearly, nonzero $\mu$ causes the growth of $X(T)$ to be curtailed, and the maximum amplitude achieved decreases with increasing $\mu$.

\subsection{Expanding background}\label{sec:expanding}

We now examine solutions of (\ref{eq:general ode}) when the expansion of the background is small but not negligible; i.e., $\Theta \ll 1$.  For simplicity, we again restrict to $n=2$ and take $\Theta$ to be constant.  To further simplify the analysis, we neglect terms proportional to $\mu\Theta$ and $\mu h$.  The equation to solve is
\begin{equation}
	\frac{d^{2}X}{dT^{2}} + 3\Theta \frac{dX}{dT} + (1 + h \cos \Omega T) X = \mu \left( \frac{dX}{dT} \right)^{2}X.	
\end{equation}
Making use of the the same ansatz as before (\ref{eq:ansatz}) and making the same approximations, we obtain the following dynamical system for the $a$ and $b$ coefficients
\begin{subequations}\label{eq:dynamical system 2}
\begin{align}
	\frac{da}{dz} & = -\frac{1}{2} (\epsilon+h) b - 3\Theta a -\frac{1}{4} \mu b(a^{2}+b^{2}), \\
	\frac{db}{dz} & = +\frac{1}{2} (\epsilon-h) a - 3\Theta b +\frac{1}{4} \mu a(a^{2}+b^{2}),
\end{align}
\end{subequations}
where
\begin{equation}
	\epsilon = \frac{\Omega^{2}-4}{2}, \quad z = \frac{\Omega T}{(9\Theta^{2}+\Omega^{2})},
\end{equation}
and we have only retained leading order terms in $\epsilon$, $h$ and $\Theta$.

We now examine the fixed points of the system (\ref{eq:dynamical system 2}).   There always exists a vacuum fixed point with $a_{0}=b_{0}=0$, which we call $\Xi$ as in the last subsection.   Linearization of the system about this fixed point yields a Jacobian matrix with eigenvalues:
\begin{equation}
	\lambda = -3\Theta \pm \frac{1}{2} \sqrt{h^{2} - \epsilon^{2}}.
\end{equation}
From this we can distinguish three principal cases when $\Theta > 0$:\footnote{We exclude the special cases $h  = \pm \epsilon$ and $6\Theta = \sqrt{h^{2}-\epsilon^{2}}$.}
\begin{enumerate}
	\item $|h| < |\epsilon|$: $\Xi$ is an attractive spiral;
	\item $|h| > |\epsilon|$ and $6\Theta < \sqrt{h^{2}-\epsilon^{2}}$: $\Xi$ is a saddle point; or
	\item $|h| > |\epsilon|$ and $6\Theta > \sqrt{h^{2}-\epsilon^{2}}$: $\Xi$ is an attractive node.
\end{enumerate}
Parametric resonance occurs only in case 2; that is, when $\Xi$ is unstable.  Also note if $\Xi$ is circle for $\Theta = 0$, it is an attractive spiral for small $\Theta$; and if it is a saddle for $\Theta = 0$, it remains a saddle for small $\Theta$.

To find the non-vacuum fixed points, it is useful to transform to polar coordinates:
\begin{equation}\label{eq:polar coords}
	a(z) = r(z) \cos \phi(z), \quad b(z) = r(z) \sin \phi(z), \quad r(z) \in \mathbb{R}^{+}, \quad \phi(z) \in [0,2\pi).
\end{equation}
In terms of these variables, the dynamical system becomes
\begin{equation}\label{eq:polar system}
	\frac{dr}{dz} = - \frac{1}{2} r (6\Theta+h\sin 2\phi), \quad \frac{d\phi}{dz} = \frac{1}{4} \left( \mu r^{2} - 2h\cos 2\phi + 2\epsilon \right).
\end{equation}
The polar coordinates of non-vaccum fixed points  $(r_{0},\phi_{0})$ must satisfy
\begin{equation}
	\sin 2\phi_{0} = -\frac{6\Theta}{h}, \quad r_{0}^{2} = -\frac{2\epsilon}{\mu} \pm \frac{2}{\mu} \sqrt{h^{2}-36\Theta^{2}}
\end{equation}
Clearly, a necessary condition for the existence of fixed points with $r_{0} \in \mathbb{R}^{+}$ is that
\begin{equation}
	|h| > 6\Theta.
\end{equation}
If $\text{sgn} (\epsilon/\mu) = -1$, this is also a sufficient condition.  We can hence conclude that there are no non-vacuum fixed points if the Hubble damping $\Theta$ is larger than $|h|/6$.

For very small damping, the non-vacuum fixed points are located at
\begin{equation}\label{eq:small Theta 1}
	\phi_{0} = \frac{1}{2} n\pi - \frac{3(-1)^{n}\Theta}{h} + \mathcal{O}(\Theta^{2}), \quad r_{0}^{2} = \frac{2[(-1)^{n}h-\epsilon]}{\mu} + \mathcal{O}(\Theta^{2}),
\end{equation}
with $n=0\ldots 3$.  Comparing to table \ref{tab:fixed points} and making use of (\ref{eq:polar coords}), we see that $n = 0,2$ corresponds to $\alpha_{\pm}$ while $n = 1,3$ corresponds to $\beta_{\pm}$.   It is obvious that, $\alpha_{\pm}$ and $\beta_{\pm}$ will be in $\mathbb{R}^{2}$ only if $r_{0}^{2}>0$.  Assuming that $r_{0}^{2}>0$ and $h>0$, it is clear that the effect of increasing $\Theta$ away from zero is to rotate $\alpha_{\pm}$ and $\beta_{\pm}$ counter-clockwise or clockwise about the origin, respectively; the reverse is true if $h < 0$.  To classify the stability of the non-vacuum fixed points, we linearize (\ref{eq:polar system}) about $(r_{0},\phi_{0})$:
\begin{equation}
	r = r_{0} + \varepsilon r_{1}, \quad \phi = \phi_{0} + \varepsilon \phi_{1}, \quad \frac{d}{dz} \left( \begin{array}{c} r_{1} \\ \phi_{1} \end{array} \right) = \left( \begin{array}{cc} 0 & -hr_{0} \cos 2\phi_{0} \\ \frac{1}{2} \mu r_{0} & -6\Theta \end{array} \right)  \left( \begin{array}{c} r_{1} \\ \phi_{1} \end{array} \right) + \mathcal{O}(\varepsilon).
\end{equation}
To leading order in the Hubble damping, the eigenvalues of the Jacobian matrix in the last equation are
\begin{equation}\label{eq:small Theta 2}
	\lambda = -3\Theta \pm r_{0} \sqrt{ \frac{(-1)^{1+n}\mu h}{2} } + \mathcal{O}(\Theta^{2}).
\end{equation}
where $n$ is odd for $\alpha_{\pm}$ and even for $\beta_{\pm}$.  If the quantity in the square root is negative, the fixed point will be a circle for $\Theta = 0$ and an attractive spiral for $0<\Theta \ll 1$.  It the quantity in the square root is positive, the fixed point will be a saddle for all $\Theta \ll 1$.

To summarize, the inclusion of a small Hubble damping term tends to take circle fixed points from the $\Theta = 0$ case and convert them into attractive spirals.  Conversely, any saddle fixed points remain as saddles for $\Theta \ll 1$.  Non-vacuum fixed points get rotated about $\Xi$ as $\Theta$ increases; the vacuum fixed point $\Xi$ remains at $(a,b)=(0,0)$.  All this means that the essential conclusion of the previous subsection remains unchanged in a slowly expanding background: non-linearities of the type induced by the models of section (\ref{sec:models}) tend to frustrate the preheating phenomenon.

\begin{figure}
\begin{center}
\includegraphics[width=\textwidth]{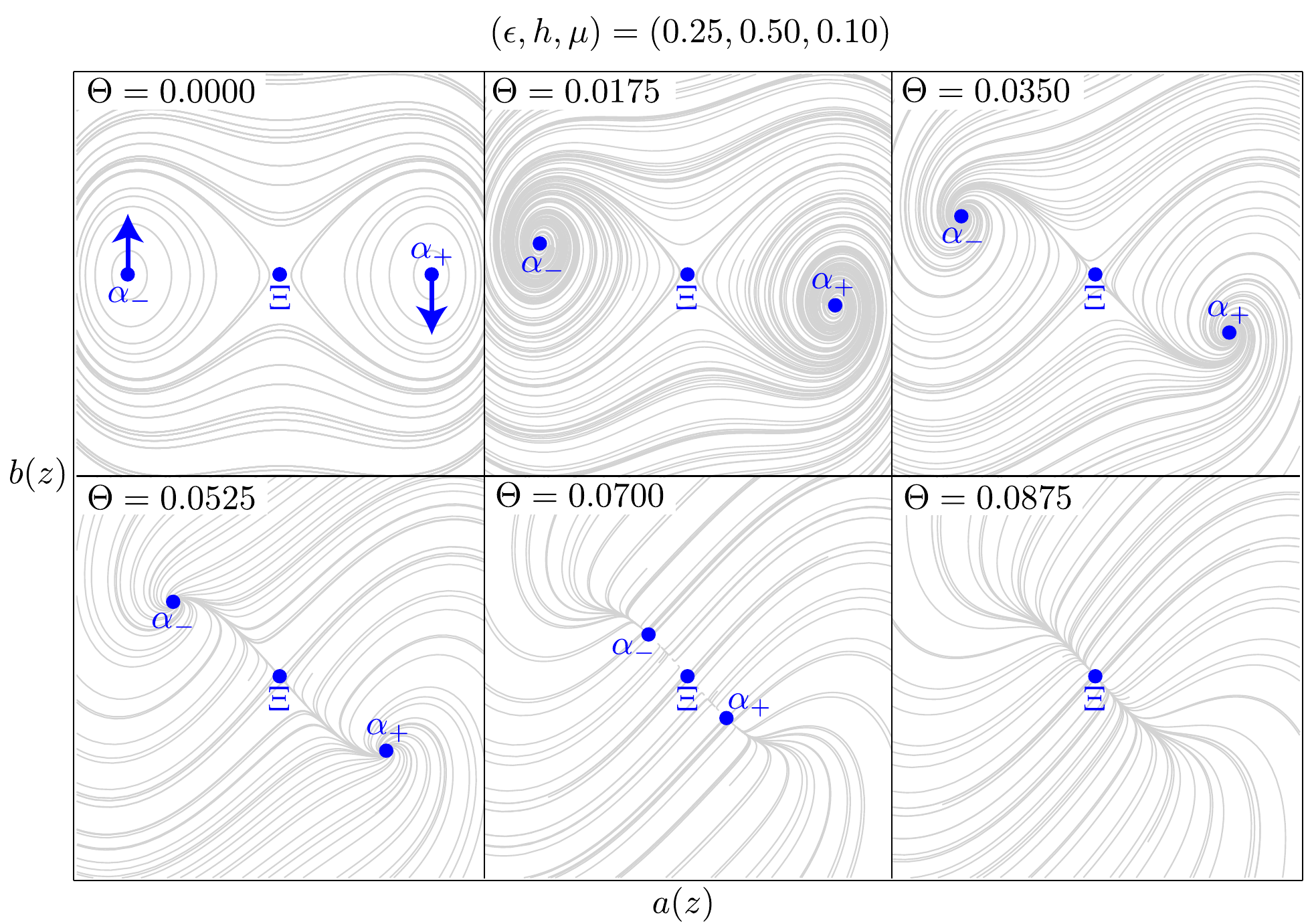}
\end{center}
\caption{A series of phase portraits showing how the dynamics associated with region IIA in the $(\epsilon,h)$ space with $\mu>0$ is affected by a non-zero Hubble parameter; i.e.\ $\Theta \ne 0$.  As $\Theta$ increases from zero, the $\alpha_{\pm}$ fixed points rotate about the vacuum saddle fixed point $\Xi$ in the clockwise direction and also switch character from circles to attractive spirals.  As $\Theta$ is further increased, $\alpha_{\pm}$ approach $\Xi$ until they eventually meet and annihilate in a pitchfork bifurcation when $\Theta = \frac{1}{6} \sqrt{h^{2}-\epsilon^{2}} \approx 0.0722$.  After the bifurcation, $\Xi$ becomes an attractive spiral.}\label{fig:expanding1}
\end{figure}
\begin{figure}
\begin{center}
\includegraphics[width=\textwidth]{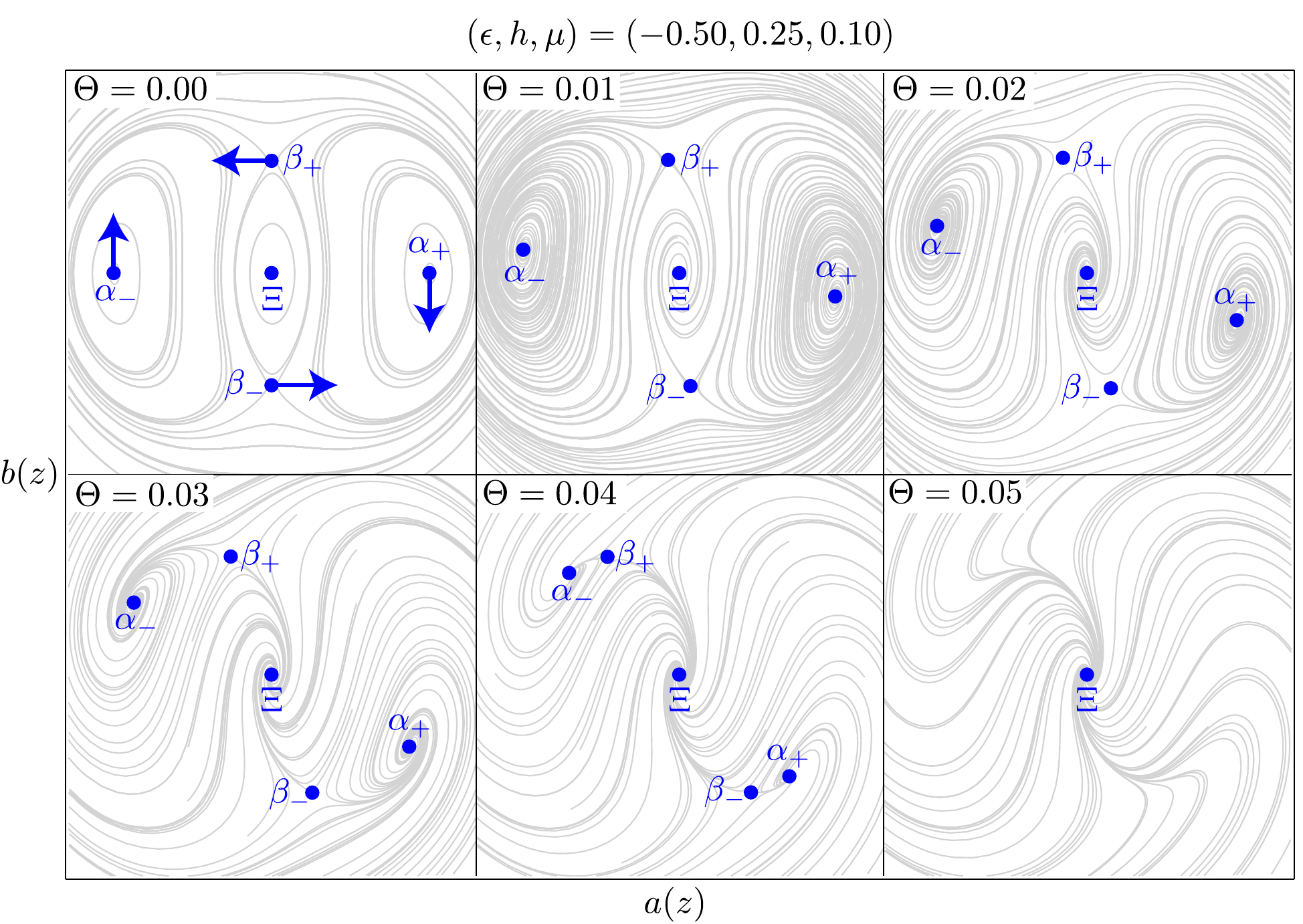}
\end{center}
\caption{A series of phase portraits showing how the dynamics associated with region IIIA in the $(\epsilon,h)$ space with $\mu>0$ is affected by a non-zero Hubble parameter; i.e.\ $\Theta \ne 0$.  As in figure \ref{fig:expanding1} the $\alpha_{\pm}$ fixed points rotate about $\Xi$ in the clockwise direction and switch from circles to attractive spirals as $\Theta$ increases from $0$.   Furthermore the $\beta_{\pm}$ fixed points rotate in the counterclockwise direction and remain as saddle points.  As $\Theta$ is further increased, the pairs $(\alpha_{-},\beta_{+})$ and $(\alpha_{+},\beta_{-})$ get closer and closer together until they eventually annihilate when $\Theta = h/6 \approx 0.042$ , leaving $\Xi$ as the sole fixed point.  $\Xi$ is an attractive spiral for all $\Theta > 0$.}\label{fig:expanding2}
\end{figure}
\begin{figure}
\begin{center}
\includegraphics[width=\textwidth]{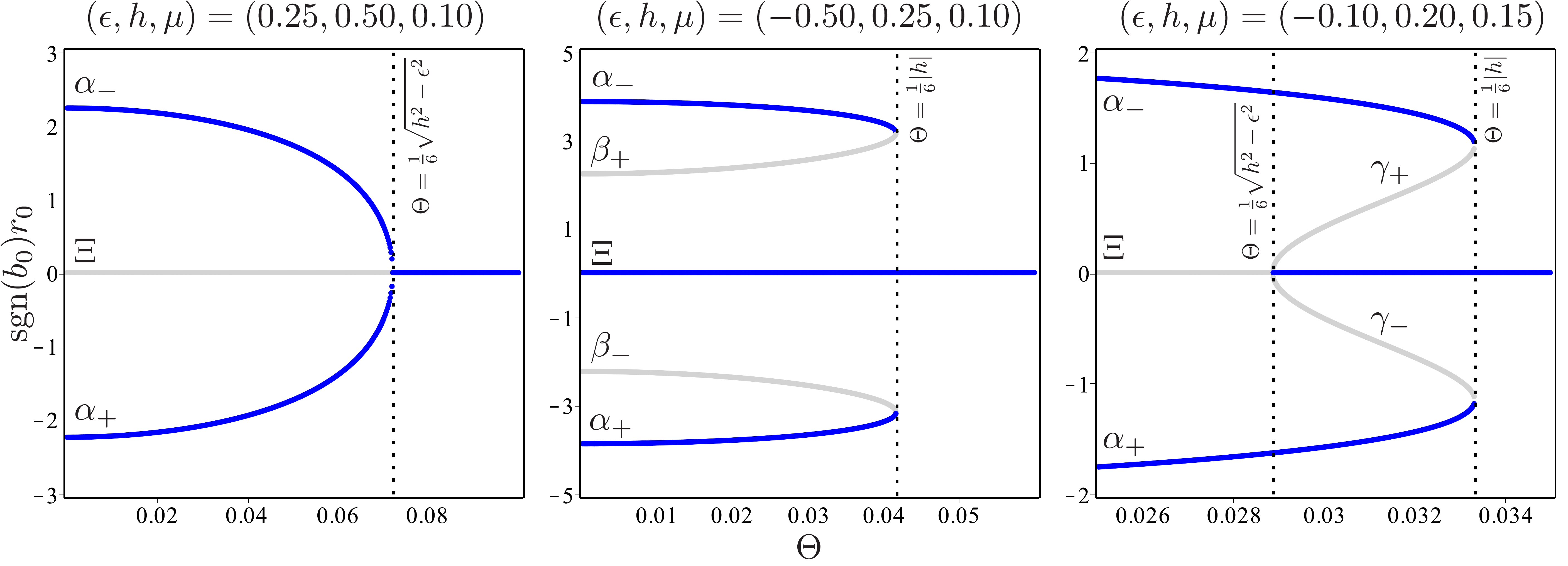}
\end{center}
\caption{Bifurcation diagrams showing the ``position'' and stability of fixed points of (\ref{eq:dynamical system 2}) as functions of $\Theta$.  Here, we characterize the location of a given fixed point (plotted on the vertical axis) by its distance from the origin $r_{0}$ times the sign of its $b$ coordinate.  If the fixed point is stable its position is indicated in dark blue, while if it is unstable the position is in light grey.  The left hand panel matches the simulations depicted in figure \ref{fig:expanding1} and demonstrates the pitchfork bifurcation where $\alpha_{\pm}$ annihilate and $\Xi$ changes stability.  The middle panel corresponds to figure \ref{fig:expanding2} and shows the annihilation of $(\alpha_{-},\beta_{+})$ and $(\alpha_{+},\beta_{-})$ in saddle node bifurcations.  The righthand panel shows a more complicated scenario where $\Xi$ first changes stability and gives birth to two new stable fixed points $\gamma_{\pm}$ in a pitchfork bifurcation.  Then at higher $\Theta$, the $\gamma_{\pm}$ annihilate with $\alpha_{\pm}$ in two saddle node bifurcations. }\label{fig:bifurcation}
\end{figure}
These analytic conclusions are based on the assumption that $\Theta$ is very small, but it is also interesting to consider the dynamics for larger values of $\Theta$ using numerical simulations.  We find that system (\ref{eq:dynamical system 2}) exhibits a rich bifurcation structure as $\Theta$ increases past the regime in which equations (\ref{eq:small Theta 1}) and (\ref{eq:small Theta 2}) are valid.  In figures \ref{fig:expanding1} and \ref{fig:expanding2}, we show how the phase diagrams of the system change as $\Theta$ is increased from 0.  As described in the captions, we see the annihilation and stability changes of fixed points as the Hubble damping changes.  In figure \ref{fig:bifurcation} we plot the corresponding bifurcation diagrams, which depict the ``position'' and stability of each fixed point as a function of $\Theta$.  Also shown in this figure is a more complicated situation involving the creation of two new fixed points $\gamma_{\pm}$ as $\Theta$ is increased.

\section{Generalization to spatially inhomogeneous reheatons}\label{sec:inhomogeneous}

A major limitation of the analysis presented so far is that we have restricted our attention to homogeneous modes.  The assumption leads to a single nonlinear ordinary differential equation (\ref{eq:general ode}) governing the resonance phenomenon that is rather simple to analyze.  However, making use of (\ref{eq:h and Theta def}), we see that the homogeneity assumption lead to the following resonance condition in the Mathieu equation ($\Omega \approx 2/N$ with $N = 1,2,3 \ldots$):
\begin{equation}
	m_\chi \approx N m_{\phi},
\end{equation}
which is a fairly stringent requirement to put on the preheating parameter space.

On the other hand, if we were to allow the field to have spatial dependence, we would have a nonlinear partial differential equation to solve.  For example, for polymer quantization the dimensionless version of the equation of motion (\ref{eq:polymer EOM}) would be:
\begin{equation}\label{eq:polymer inhomogeneous}
	\frac{\di^{2}X}{\di T^{2}}  - D^{2}X +  (1 + h \cos \Omega T)X = \mu \left( \frac{\di X}{\di T} \right)^{2} (X-D^{2}X),	
\end{equation}
where we have neglected the cosmic expansion ($\Theta = 0$) and expanded to linear order in $(h,\mu)$.  The $D$ operator is a dimensionless version of the spatial Laplacian
\begin{equation}
	D^{2} =  \omega_{0}^{-2} \nabla^{2}.
\end{equation}
Of course, if $\mu = 0$ this can be solved in terms of eigenfunctions of $D^{2}$; i.e., Fourier modes:
\begin{equation}\label{eq:Fourier decomp}
	X(T,\mathbf{Z}) = \int d\k \,  Q_{\k} (\mathbf{Z}) f_{\k}(T), \quad D^{2} Q_{\k} = -\frac{k^{2}}{\omega_{0}^{2}} Q_{\k},
\end{equation}
with $\mathbf{Z} = \omega_{0} \mathbf{x}$.  The Fourier amplitudes $f_{\k}$ are time dependent and satisfy:
\begin{equation}\label{eq:Fourier EOM}
	\frac{d^{2}f_{\k}}{d T^{2}}  +  \left( \frac{k^{2}}{\omega_{0}^{2}}  + 1 + h \cos \Omega T \right)f_{\k} = 0.	
\end{equation}
This is again the Mathieu equation, and using (\ref{eq:dimensionless}) and (\ref{eq:h and Theta def}) the condition for parametric resonance is
\begin{equation}
	k^{2} \approx N^{2} m_{\phi}^{2}-m_{\chi}^{2}.
\end{equation}
This means that no matter what the relative sizes of $m_{\phi}$ and $m_{\chi}$ are, there will alway an infinite number of Fourier modes with $k \approx \sqrt{N^{2} m_{\phi}^{2}-m_{\chi}^{2}}$ prone to exponential growth via parametric resonance.  In this way, one can argue in the linear $\mu =0$ case that parametric resonance is a rather generic feature.

\begin{figure}
\begin{center}
\includegraphics[width=\textwidth]{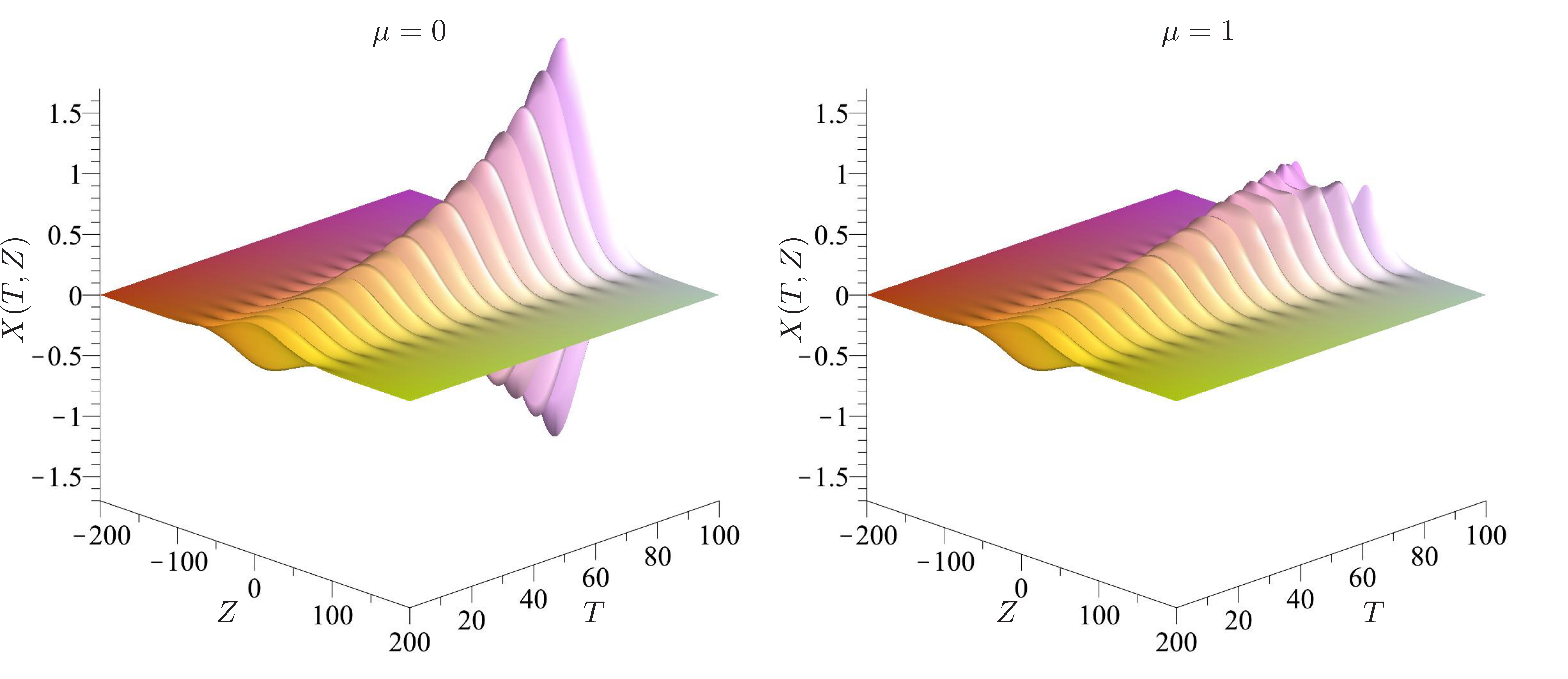}
\end{center}
\caption{Simulations of parametric resonance in $1+1$ dimensions as governed by equation (\ref{eq:polymer inhomogeneous 1+1}).  We present results for standard preheating (\emph{left}) and when the matter field equation of motion is corrected by polymer quantization (\emph{right}).  We see that the resonance phenomenon is effectively shut down by the $\mu \ne 0$ nonlinearity for $T \gtrsim 80$.  Also observe the growth of small scale structure in the nonlinear simulation for late times.}\label{fig:1+1}
\end{figure}
But when $\mu \ne 0$, one cannot perform the separation of variables (\ref{eq:Fourier decomp}) to obtain the linear equation (\ref{eq:Fourier EOM}); rather, the nonlinearity on the righthand side of (\ref{eq:polymer inhomogeneous}) induces non-trivial coupling between Fourier modes.  It is not clear if the frustration of parametric resonance that we found for $k=0$ modes will persist to the $k\ne 0$ case.  To investigate this, we have performed numeric simulations of (\ref{eq:polymer inhomogeneous}) in a toy model with one spatial dimension; i.e., we solve the equation
\begin{equation}\label{eq:polymer inhomogeneous 1+1}
	\frac{\di^{2}X}{\di T^{2}}  - \frac{\di^{2}X}{\di Z^{2}} +  (1 + h \cos \Omega T)X = \mu \left( \frac{\di X}{\di T} \right)^{2} \left( X-\frac{\di^{2}X}{\di Z^{2}} \right).
\end{equation}
When performing simulations, we assume initial data with compact support.  For a simulation that runs over the interval $T \in [0,T_{0}]$, we take the spatial domain to be $X \in [-L,L]$ where $L>T_{0}$ is selected to ensure that the boundaries are outside the causal future of the initial data.  Typical simulation results are shown in figure \ref{fig:1+1} for $\mu=0$ and $\mu=1$.  In the standard $\mu=0$ case, we see the exponential growth of the initial profile as expected.  The first part of the $\mu=1$ simulation also shows exponential growth, but this eventually levels off after some finite value of $T$.  This tends to support the hypothesis that our results obtained for the homogeneous case should generalized to inhomogeneous preheating; i.e., the nonlinearity effectively shuts down parametric resonance,

However, there are a couple important caveats that should be stated: Perhaps most obviously, we only simulate one spatial dimension here, so there is some uncertainly on how these results will generalize to higher dimensions.   Second, it is time consuming to run simulations long enough to resolve an periodic variations in amplitude like the ones seen in figure \ref{fig:simulations}.  This is because the non-linearity in (\ref{eq:polymer inhomogeneous 1+1}) tends to create smaller and smaller structures in the solution as time goes on.  This is easy to see perturbatively: if one assumes a small amplitude zeroth order solution $X(T,Z) \propto e^{ikZ/\omega_{0}}$ and then feeds it back into the righthand side of (\ref{eq:polymer inhomogeneous 1+1}) to find corrections proportional to $\mu$, one immediately sees that such corrections involve terms $\propto e^{3ikZ/\omega_{0}}$.  That is, the nonlinearity will cause modes with 1/3 the size of the initial data to grow.  As the simulation proceeds, these modes will induce other modes with $1/9$ the initial data size to increase in amplitude, and so on.  One can actually see this in figure \ref{fig:1+1}:  When $\mu = 0$, the spatial profile of $X(T,Z)$ remains more or less the same while the amplitude increases.  Conversely, when $\mu = 1$ we can see the formation of non-trivial small scale structure in the later stages of the simulation.  The transfer of power from large scales to small scales in nonlinear partial differential equations is known as ``weak turbulence'', and it represents a significant challenge for numerical simulations.  Hence, we cannot say with certainty that $X(T,Z)$ remains bounded at $T \rightarrow \infty$, it is possible that small scale structures in the solution grow exponentially as time progresses.

In summary, the numeric simulations we present in this section suggest that inhomogeneous parametric resonance will also be frustrated by the types non-standard kinetic behaviour considered in this paper, but there is enough uncertainty in this hypothesis to warrant more detailed studies in the future.

\section{Conclusions}\label{sec:conclusions}

In this paper, we have studied the effects of non-standard kinetic terms on the parametric resonance phenomenon in the preheating period after inflation.  We considered a number of different models with somewhat diverse motivations, and found that to leading order they each led to very similar reheaton equations of motion, generally involving cubic nonlinear self-interactions.  In the spatially homogeneous limit and neglecting cosmological expansion, we studied the nonlinear reheaton dynamics for parameters close to resonance using a multiple scales analysis.  We concluded analytically and confirmed numerically that the self-interaction shuts down the parametric resonance effect after a finite amount of amplification of the reheaton amplitude.  We then considered the effects of a small Hubble parameter and found that the governing dynamical system exhibits a rich bifurcation structure.  However, the main conclusion that long-wavelength parametric resonance is frustrated is unaffected by cosmological expansion.  Finally, we examined the behaviour of a spatially inhomogeneous reheaton subject to polymer quantization in $(1+1)$--dimensions.  We found numeric evidence that the frustration of parametric resonance persists if we drop the long-wavelength assumption, but our results should not be construed to be definitive.

At least one direction for future work is obvious:  to determine whether or not the effects found in this paper are robust, one would need to perform full $(3+1)$--dimensional simulations in the same vein as reference \cite{Child:2013ria}, except with non-minimal kinetic terms applied to the reheaton instead of the inflaton.   From these, one could determine to what extent exotic nonlinearities in the matter sector really do limit the effectiveness of resonant preheating in the early universe.

\acknowledgments
We would like to thank Saurya Das, Bret Underwood and James Watmough for useful discussions.  We are supported by National Sciences and Engineering Research Council of Canada (NSERC).

\bibliographystyle{apsrev4-1}
\bibliography{preheating}

\end{document}